\documentclass[epsfig,12pt]{article}
\usepackage{graphicx}
\usepackage{amssymb}
\usepackage{setspace}
\begin{document}

\begin{center}
\textbf{Understanding the complexity of the L\'evy-walk nature of human mobility with a multi-scale cost/benefit model}\\[0pt]

Nicola Scafetta$^{1}$ \\[0pt]
\end{center}

$^1$ ACRIM and Duke University, Durham, NC 27708.

%%%%%%%%%%%%%%%%%%%%%%%%
%%%%%%%%%%%%%%%%%%%%%%%%%%%%%%%%

ABSTRACT:

Probability distributions of human displacements has been fit with exponentially truncated L\'evy flights or fat tailed Pareto inverse power law probability distributions. Thus, people usually stay within a given location (for example, the city of residence), but with a non-vanishing frequency they visit nearby or far locations too.  Herein, we show that an important empirical distribution of human displacements (range:  from 1 to 1000 km) can be well fit by three consecutive Pareto distributions with simple integer exponents equal to 1, 2 and ($\gtrapprox$) 3. These three exponents correspond to three displacement range zones of about 1 km  $\lesssim \Delta r \lesssim$ 10 km,   10 km $\lesssim  \Delta r \lesssim$ 300 km and  300 km $\lesssim \Delta r \lesssim $ 1000 km, respectively. These three zones can be geographically and physically well determined as displacements within a city, visits to nearby cities that may occur within just one-day trips, and visit to far locations that may require multi-days trips.  The incremental integer values of the three exponents can be easily explained with a three-scale mobility cost/benefit model for human displacements based on simple geometrical constrains. Essentially, people would divide the space into three major regions (close, medium and far distances) and would assume that  the travel benefits are randomly/uniformly distributed mostly only within specific urban-like areas. The three displacement distribution zones appear to be characterized by an integer (1, 2 or $\gtrapprox$3) inverse power exponent because of the specific number (1, 2 or $\gtrapprox$3) of cost mechanisms (each of which is proportional to the displacement length. The distributions in the first two zones would be associated to  Pareto distributions with exponent $\beta=1$ and $\beta=2$ because of simple geometrical statistical considerations due to the a-priory assumptions that most benefits are searched in the urban area of the city of residence or in specific nearby cities.   We also show, by using independent records of human mobility, that the proposed model predicts the statistical properties of human mobility below 1 km ranges, where people just walk. In the latter case, the threshold between zone 1 and zone 2 may be around 100-200 meters and, perhaps, may have been evolutionary determined by the natural human high resolution visual range which characterize an area of interest where the benefits are assumed to be randomly and uniformly distributed.  This rich and suggestive interpretation of human mobility may characterize other complex random walk phenomena that may also be described by a N-piece fit Pareto distributions with increasing integer exponents. This study also suggests that distribution functions used to fit experimental probability distributions must be carefully chosen for not improperly obscure the physics underlying a physical phenomenon.

\newpage

\textbf{
Human motion is a particular type of recurrent diffusion process where agents move from a location to another and return home after a given complex trajectory made of a certain number of steps of given lengths. For a large set of agents this motion can be considered equivalent to a complex random walk described by a peculiar statistics. Understanding the statistical nature of human motion is the topic of the present work. In the literature, human displacement distributions are traditionally fit using an exponentially truncated Pareto or inverse power law distribution based on a single inverse power-law exponent. However, the physical meaning of this specific distribution function, as well as of the measured inverse power law exponent (for example, $\beta= 1.75$) have been never explained. Herein we show that the same human displacement distribution can be better fit with three inverse power law distributions described by simple integer exponents ($\beta$ = 1, 2, 3 or larger). Each exponent characterizes a specific displacement length zone whose ranges are about 1 km $\lesssim \Delta r \lesssim$ 10 km,   10 km $\lesssim  \Delta r \lesssim$ 300 km and  300 km $\lesssim \Delta r \lesssim $ 1000, respectively. This alternative fit methodology for interpreting human displacement distributions is crucial because it suggests a clear physical/geometrical generating mechanism. Essentially, people would  divide the space into three regions (close, medium and far distances), and they would assume that in each zone region the travel benefits  are randomly/uniformly  distributed mostly within the urban area of the cities and would associate to each zone region an increasing integer number of cost mechanisms required to cover the given distance. We suggest that the probability associated to a displacement length is a monotonic function of its \emph{cost}: less expensive (or shorter) displacements take place more likely than more expensive (or longer) displacements. Human mobility may be conditioned by multiple independent cost functions that work alone or may be statistically combined together according to the displacement length. For example, typical cost functions taken into account by people are the cost of the fuel, the time duration for the displacement, lodging costs that are required for very long trips, and others. The crucial fact is that on average these cost functions are likely directly proportional to the length of the displacement itself. For example, on average the fuel cost for covering a distance $L$ is half the cost for covering a distance $2L$: the same is true for the time interval needed to cover a give distance. Thus, we can expect that the simplest strategy adopted by humans in optimizing their movements is that when fuel cost alone are taken into account by a person, displacements of length $L$ are twice more probable than displacements of length $2L$ because of the homogeneous benefit distribution assumption within each urban zone. Thus, by assuming that people optimize their movement by uniformly distributing their travel resources (in time, energy and money) within given range zones, we demonstrate by simple geometrical considerations based on area ratios that each displacement cost function would yield  to a displacement probability distribution described by a basic inverse power law distribution $P(\Delta r)\propto 1/ \Delta r$. When nearby cities are visited a similar reasoning would yield to inverse power law with $\beta=2$ because at leat 2 cost basic mechanisms would be activated and because the searched benefit would still be mostly searched within an urban zone. If 3 or more displacement cost functions condition the decision of an agent as for visits to far locations, then the combined probability distribution is the product of more basic probability distributions and would be characterized by inverse power law distributions with integer exponent $\beta$ = 3 or more. We also show that a similar model predicts the statistical properties of human displacements below 1 km ranges by people who just walk, and whose walking decision could be determined by time and physical energy considerations. In the case of just walking people the threshold between zone 1 and zone 2 may be a few hundred meters and, perhaps, it may have been evolutionary determined by the natural human high resolution visual range. The peculiarity of the three-scale pattern shown by the human displacement distribution suggests that on average humans take into account one, two, three (or more) alternative displacement cost mechanisms according to the length of the displacement itself and assume that the benefits are   uniformly/randomly located only within specific areas of interests, which characterize zone 1. This property may be more general and may characterize other complex random walk phenomena that are commonly interpreted in the scientific literature with simple exponentially truncated inverse power law distributions based on a single exponent that may obscure, instead of clarifying, the physical mechanism behind a physical phenomenon.}

\section{Introduction}

Complex random walks (from Brownian fractional diffusion to L\'evy walk and flights) are interesting physical phenomena that have found several applications in physics \cite{Scafetta2010} including human gait dynamics \cite{Scafetta}. One of the most intriguing complex motion is the human one.
Understanding human mobility laws  is important for urban planning \cite{Horner}, traffic forecasting \cite{Kitamura1,Kitamura2}, and  for understanding the spread of biological \cite{Colizza,Eubank,Hufnagel} and mobile viruses \cite{Kleinberg}. Human mobility also determines the formation of several social networks. In fact, in several cases the formation of human networks necessities the physical displacements of individuals. The social environment where humans move and make encounters,  forming links and consequently networks, is mostly made of the residence city   where people mostly live and work, and by a set of nearby or far locations and cities that can be occasionally visited. Thus, understanding human mobility is also important for determining how specific social networks form and grow. Small-world social networks \cite{Schnettler} as well as more complex networks such as free scale-social networks and others \cite{Toivonen} may be partially determined and explained by the specific patterns that characterize human mobility.

Herein, we propose a specific model for understanding human mobility that takes into account the fact that a careful analysis of empirical data referring to human displacements reveals that this phenomenon may  be characterized by a  multiple-scale L\'evy flight cost mechanism model constrained by simple spatial and temporal considerations that should constrain human mobility.

The purpose of our study is to propose a possible physical explanation of human mobility interpreted in its overall statistical properties as phenomenologically manifested in probability distributions obtained from data collected from a very large sample of people. Thus, our study is not aimed to determine the dynamics of the mobility of each person which, of course, will be very different from person to person and from location to location. However, when the dynamics of a very large sample of persons  in a multitude of  locations is averaged in an overall probability distribution of displacements, the individual properties of each trajectory are smoother out and the emerging statistical pattern would manifest a common underling characteristic of human mobility, which is the topic addressed in the present study.

Empirical distributions referring to human and animal displacements have been often approximated with various random walk or diffusion models that typically obey Pareto inverse power law or  L\'evy  flight distributions \cite{Gonzalez,Brockmann,Viswanathan,Edwards,Ramos,Sims,Rhee}. Inverse power law distributions are characterized by  \emph{fat} probability distribution tails. These tails indicate that while most human displacements are relatively short, the frequency of long displacements is not negligible either. Trips to nearby or far locations are rare, but not statistically impossible! If on average six links are sufficient to connect everybody to everybody else \cite{Watts}, one explanation may be because of the peculiar inverse power law distribution that characterizes human mobility. In fact, the fat   tail of these probability distributions implies that people can meet and form links also with individuals living quite far from their residence location.

Pareto or inverse power law distributions are characterized by a specific power exponent $\beta$ such as:
\begin{equation}\label{eq66}
    P(x)\propto \frac{1}{x^\beta}.
\end{equation}
Numerous physical, biological and social systems are characterized by Pareto distributions \cite{Schroeder}. However, there are a few fundamental methodological and physical problems that need to be addressed for properly understanding the physics underlying a given complex phenomenon described by a probability distribution.

1) Eq. (\ref{eq66}) is a mathematical abstraction. No real system can be described by such a function for any real value of the variable $x$. In real systems the tail of the distribution is truncated because all natural systems are physically limited.

2) Real complex systems are often not described by just one scaling exponent.  Real complex systems are often multi-scales, that is, they are characterized by probability distributions that can be divided into different range zones. Each distribution zone may be characterized by a set of physical properties that determine the specific geometry of its probability distribution within such a zone. Examples of bi- or multi-mode distributions are common \cite{Latapy}.

3) Finally, the specific value of the measured exponents of these inverse power law distributions does have a physical origin that has to be properly investigated and explained.

Probability distributions can also be fit with alternative functions generated by very different physical background mechanisms, as the super-statistics of Beck and Cohen \cite{Beck} has shown. In general, it may be even difficult to correctly determine which distribution function should be used to properly fit and interpret a given complex probability distribution.

For example, some authors \cite{Fenner} have adopted exponentially truncated inverse power law function. Gonzalez \emph{et al.} \cite{Gonzalez} and Edwards  \emph{et al.}  \cite{Edwards} explicitly fit their human displacement distributions with an exponentially truncated inverse power law distribution because Eq. (\ref{eq66}) would not fit the data. The same is done by Rhee \emph{et al.} \cite{Rhee} where their displacement distributions obtained from people who mostly walk are fit with truncated Pareto distributions. The choice of these authors, however, is not fully backed by a clear physical model and was proposed without a single explanation.

Indeed, an exponential decay of the tail of a displacement distribution may be due to poor statistics or by geographical constrains that would not allow animals or humans to move for a distance longer than a certain maximum. For example, it is reasonable to assume that the L\'evy  flight patterns observed in wandering albatrosses  may terminate with an exponential cutoff because these flights may be as long as several thousand kilometers and this geographical scale  becomes compatible with the size of the Earth \cite{Viswanathan}. However, the issue may be more complicated and specific geometries of the displacement distribution    may also indicate specific physical characteristics of human/animal movements that should be properly investigated.

The purpose of this work is to study whether human displacement distributions present more complex patterns that cannot be properly described by an exponentially truncated inverse power law distribution. In particular, we will show evidences that human displacement data can be described by a multi-scale inverse power law distribution that would eventually approach an exponential cut off. Then, we will propose that this multi-scale inverse power low distribution can be explained using a multi-scale statistical displacement cost mechanism grounded on commonsense physical and geometrical considerations that should characterize human mobility.

\section{The traditional fit methodology: range from 1 km to 1000 km}

One of the first large scale human motion studies was attempted by  Brockmann \emph{et al.} \cite{Brockmann}. These authors based their analysis of human movements on the trajectories of 464,670 dollar bills obtained from the bill-tracking system. From those reports these authors calculated the geographical displacements $\Delta r  = |x_A-x_B|$ between a first ($x_A$) and secondary ($x_B$) report location of a bank note and the elapsed time $\Delta T$ between successive reports. They concluded that the distribution of these displacements obeys an inverse power law $P(\Delta r) \propto \Delta r ^{-\beta}$  with  exponent  $\beta = 1.59$. However, Brockmann \emph{et al.}'s data  do not allow to exactly track human movements, but, more properly, bank note movements that pass from hand to hand. The observed distribution does not reflect the motion of individual users but some kind of convolution between population-based heterogeneities and individual human trajectories.

More recently, Gonzalez \emph{et al.} \cite{Gonzalez} based their large scale study on trajectories of 100,000 anonymized mobile phone users. Each time a user initiated or received a call or a text message, the location of the mobile phone tower routeing the communication was recorded, allowing to reconstruct both user's time and space resolved trajectory within a reasonable resolution. In this way, these authors could measure the distance between user's positions at consecutive calls. They collected a very large data set:  16,264,308 displacements over a six-month period. This data set was labeled as `D1' in Gonzalez \emph{et al.}.  Because of its large size, this data set is probably the best available record  for attempting to understand whether some simple statistical law governs human mobility. Herein we study the displacement distribution proposed by Gonzalez \emph{et al.} that is depicted in Figure 1.

 The area covered by each mobile phone tower has a varying extension. Inside urban areas the distances between adjacent towers is lower than 1 km. In the rural regions   the mobile phone towers become sparser and their relative distance is up to a few kilometers. Thus, the distribution depicted in Figure 1 represents human mobility with the minimum resolution of about 1 kilometer. The distribution range falls within 1000 km. This range is well below the geographical size of Europe where the data were collected from a cellphone provider. Thus, the entire  range of the distribution for 1 km $<\Delta r <$ 1000 km, or at least the range from 1 km to 300 km, likely reflects a real property of human mobility and should be explained by a physical model.

 Gonzalez \emph{et al.} \cite{Gonzalez} fit in their distribution (depicted in their figure 1C) with an exponentially truncated inverse power law (L\'evy  flight) function of the type:
\begin{equation}\label{eq1}
    P(\Delta r) \propto (\Delta r + \Delta r_0)^{-\beta} \exp(-\Delta r/K).
\end{equation}
The fit gives:  $\beta= 1.75 \pm  0.15$,  $\Delta r_0 = 1.5$ km and cutoff values $K \approx 400$ km. Analysis of the data  also show that individual users return to a few highly frequented locations with a daily frequency, such as home and work office, and there exists a weekly recurrence cycle as everybody would expect that humans do. This recurrent movement is quite different from, for example, bank notes that, instead, diffuse \cite{Brockmann}. Thus, Gonzalez's record is more appropriate for studying human mobility.

However, no physical explanation has been proposed by Gonzalez \emph{et al.} to justify Eq. \ref{eq1}. It appears that the function (\ref{eq1}) has been chosen to fit the data just because this is how this kind of distributions is traditionally fit in the literature, and because this function can approximately fit the data with appropriate regression parameters. No physical explanation of the found exponent $\beta=1.75$ has been proposed by Gonzalez \emph{et al.}. Thus, no real physical model has been proposed to explain the distribution depicted in Figure 1.

\section{An alternative fit methodology}

The present work focuses on an alternative interpretation of the distribution depicted in Figure 1. In fact,
 Figure 2 shows that the  distribution of human displacement can be divided into  three different zone ranges that can be fit with three inverse power laws of the type of  Eq. (\ref{eq66}): $p_\beta (r)\propto 1/\Delta r^\beta$.  We get: Zone 1, $\beta  \approx 1$ for  $1<\Delta r < 10$ km;  Zone 2,  $\beta  \approx 2$ for  10 km $<\Delta r < $ 300 km; Zone 3,   $\beta  \approx 3$ for  300 km $<\Delta r < 1000$ km. The distribution is assumed to be statistically meaningful  within all represented ranges from about 1 km to 1000 km, as claimed by Gonzalez \emph{et al.} \cite{Gonzalez}.

From a purely statistical perspective fitting the distribution with multiple inverse power law functions as done in Figure 2 appears to be statistically more accurate than fitting it as done in Figure 1. This would be true at least within the range from 1 km to 300 km. In fact, the average $\chi^2$ between the logarithm of the data and the fitting function model predictions gives: in the interval [1 km: 10 km], $\overline{\chi^2}=0.0027$ with Eq. (\ref{eq1}) while $\overline{\chi^2}=0.0020$ with Eq. (\ref{eq66}) and $\beta=1$; in the interval [10 km: 300 km], $\overline{\chi^2}=0.0039$ with Eq. (\ref{eq1}) while $\overline{\chi^2}=0.0021$ with Eq. (\ref{eq66}) and $\beta=2$; in the interval [300 km: 1000 km], $\overline{\chi^2}=0.020$ with Eq. (\ref{eq1}) while $\overline{\chi^2}=0.026$ with Eq. (\ref{eq66}) and $\beta=3$.

Thus, for displacements larger than 300 km, Eq. (\ref{eq1}) performs slightly better than Eq. (\ref{eq66}), but this data are statistically poorer than those covering the range from 1 km to 300 km, in particular at the very end of the distribution. Consequently, it may be possible the data may follow a Pareto law (\ref{eq66}) with $\beta=3$, as well as they may present an exponential  cut off or, alternatively, they may be described with an inverse power law with a larger integer exponent. However, for displacements from 1 km to 300 km, where the data are more numerous and the distribution is statistically more robust, the distribution is clearly better  fit with two consecutive inverse power law distributions with $\beta=1$ and  $\beta=2$, which refer to two different displacement ranges.

At a first sight the difference between the two alternative ways (depicted in Figure 1 and 2, respectively) of fitting the distribution of human displacements appears to be so small that most people would consider the entire exercise to be irrelevant. However, a vigilant reader would recognize that the two alternative fitting methodologies are physically quite different. In the following section we show that the latter method (depicted in Figure 2) suggests an evident physical model of human mobility, while the former one (depicted in Figure 1) does not have any evident physical meaning. Thus, our exercise is not meaningless. It is a step forward a physical understanding of  the complexity of the L\'evy-walk nature of human mobility based on a careful analysis of the data, which  a simple exponentially truncated inverse power law distribution function with a single exponent might have obscured.

\section{A simple multi-scale diffusion model of human displacements and its statistical/physical interpretation}

The interpretation of the above finding takes into account that geometrical constrains and travel costs both in terms of time and money determine the frequency of any kind of human displacement \cite{Lundevaller}. This behavior appears to be a general economical characteristic. In fact,  in Krugman's economic geography \cite{Fujita}, it is argued that people tend to cluster in cities (instead of distributing themselves homogeneously on the planet) partly because it saves on the transportation costs both in terms of time and money. The clustering of the benefits in restricted areas produces geometrical constrains that force the distributions of displacements to assume an inverse power law distributions characterized by different integer exponents according to whether an agent is moving within an area or between area regions.

I fact, according to the results depicted in Figure 2, the distribution of human displacements can be approximately separated into three consecutive displacement zones (relative to short, medium and long displacements).   Each zone can be fit with three inverse power law functions with exponents approximately equal to 1, 2 and ($\gtrapprox$)3, respectively. This result suggests that human mobility may have a simple statistical explanation that can be based on simple displacement cost functions \cite{Lundevaller} and geographical topology.

The almost sudden statistical transition observed from a displacement zone to the following one suggests that the number of cost functions involved in the process increases with the number that characterizes the displacement zone (1, 2 or 3). Thus, we propose a statistical model based on the empirical finding itself shown in figure 2 that assumes that the displacement probability distribution of human mobility is given by
\begin{equation}\label{eq2}
    P(\Delta r)\propto \prod_{i=1}^{N(\Delta r)} \frac{1}{f_i(\Delta r)},
\end{equation}
where the functions $1/f_i(\Delta r)$ are the probability functions associated to independent cost mechanisms that a given displacement  $\Delta r$ requires. Note that taking into account more than one cost mechanism should introduce a multiplicative term in the probability distribution because in statistics if $A$ and $B$ are two independent events the probability of their union is equal to the product of the two probabilities:  $P(A \cup B)=P(A)P(B)$. In our case it is evident that if two or more independent cost mechanisms are taken into account by an agent, the overall probability would be a weighted function of all single probabilities associated to each cost function: that is, the overall probability  is given by the product of the single probability functions associated to each cost mechanism, as well known in statistics.

The peculiarity of the model is that the number of cost mechanisms involved in a displacement increases with the displacement zone number. Thus, we have: $N(\Delta r) = 1$  for $\Delta r<\Delta r_1$; $N(\Delta r) = 2$  for $\Delta r_1<\Delta r<\Delta r_2$; $N(\Delta r) = 3$  for $\Delta r_2<\Delta r<\Delta r_3$, and potentially up to N zones. The displacements $\Delta r_i$ with $i=1,2,3,\dots, N$ are the displacement borders that separate the  zones. The model can be easily extended to N zones.

Eq. \ref{eq2} also implies that
\begin{equation}\label{eq3}
    P(\Delta r) \prod_{i=1}^{N(\Delta r)}f_i(\Delta r) = C_{N(\Delta r)},
\end{equation}
where $C_j$ with $j=1,2,3,\dots$ are specific constants that characterize each displacement zone. The constants implicit in Eq. \ref{eq3} would imply that a typical individual first divides the entire space around him in three major zones (close, medium and far distances). In fact, classifying things into three groups by size (small, medium, and large) is a quite natural methodology, as also Aristotle would teach us.

Eq. \ref{eq3} may also indicate that from a statistical perspective people optimize their movement by homogeneously distributing their resources in the geographic space available. This also would imply a travel length frequency inversely proportional to the overall (money and/or time) travel cost, which would yield to Eq. \ref{eq2}.

Costs that people take into account are travel time, fuel and lodging expenses, which are constrained by money budgets \cite{Mokhtarian}.
The crucial observation is that travel fuel cost and time duration are approximately proportional to the displacement length itself. Moreover,  travel fuel cost and time duration are physically independent and different cost functions.   Thus, for example, we can assume that if a person takes into account only the cost of the fuel or only the time duration  for moving from a location A to a location B, a displacement of a given length $\Delta r$  will be twice as more probable than a displacement of a given length $2\Delta r$. This property, together with the assumption that the benefits are randomly and uniformly distributed within an urban area, should  yield, for each cost mechanism, to  Pareto distributions with exponent $\beta=1$.

The proposed hypothesis  that the probability distribution associated to a single cost function is given by a Pareto distributions of the displacement length with exponent $\beta=1$ should be understood as a direct consequence of the observed L\'evy nature of the distribution of human displacements, as implicit in the probability distribution  depicted  in Figure 2. In fact, it is the L\'evy nature of the distribution of human displacements itself that appears to exclude other possible distribution functions of the displacement such as, for example, an uniform distribution or an exponential distribution or a Poisson distribution or others, which would not be characterized by inverse power law L\'evy's tails.

A simple statistical explanation of why within an urban area the probability associated to a cost mechanism determining a given displacement may be well described by Pareto distributions of the displacement length with exponent $\beta=1$ can be easily argued on geometrical considerations alone that would constrain the statistics by favoring the emergence of such distributions. In fact, if the benefits are assumed to be uniformly/randomly distributed within an area of interest such as within a city, one of the reasons why a person may decide to move by a certain distance $\Delta r=R$ is because s/he cannot find the desired benefit within a shorter distance $\Delta r<R$. Thus, two independent events must simultaneously occur: event A) the desired benefit is not found  within the distance $\Delta r<R$: event B) the desired benefit is found at the distance $\Delta r=R$. Moreover, because we are assuming that within an urban area the benefits are uniformly/randomly distributed on a surface, the probability goes with the covered area; the probability of \emph{not} finding a desired benefit within the distance $\Delta r<R$ must decrease with the area of a circle of radius $R$ and should be proportional to $R^{-2}$, while the probability of finding a benefit at a distance $\Delta r=R$ must increase with the area of a ring of radius $R$ and should be proportional to $R$ itself. Thus, the combined probability between event A and event B is given by the following equation:

\begin{equation}\label{pron}
    P(A \cup B)=P(A)P(B)\propto \frac{1}{R^2} \times R \propto \frac{1}{R},
\end{equation}
which is a Pareto distributions of the displacement length with exponent $\beta=1$. In conclusion, we suggest that the statistics associated to the above geometry forces the displacement distribution of human movement, which should be directly related to the displacement cost, to adopt an inverse power law formulation with exponent  $\beta=1$. Because this is the simplest case, we can assume that when a single cost mechanism is taken into account, it naturally produces Pareto displacement histograms with exponent $\beta=1$ as constrained by geometrical properties.

Zone 1 refers to displacements from about 1 km to about 10 km that mostly refers to movements inside a city. The city where a individual is located is likely felt as a \emph{close} zone. Within this range Figure 2 shows that the probability distribution is approximately proportional to $1/\Delta r$. In fact, within an urban zone, a person may be  directly influenced by the geometrical statistical constrains expressed in Eq. \ref{pron} and s/he may take into account only one cost mechanism. For example, the decision of moving by a certain distance $\Delta r$ may be conditioned by fuel cost considerations or by time duration considerations, but not by both. Thus, according to this interpretation, within the range of a few kilometers, each person thinks about saving fuel or, alternatively, saving time, but on average only one cost mechanism proportional to the displacement distance $\Delta r$ is taken into account by each person for each displacement decision, and the distribution associated to this cost would be constrained by the geometrical property expressed by Eq.  \ref{pron}.
So, for short displacements we have that their frequency is inversely proportional to the travel cost (time duration or fuel consumption) which is approximately proportional to the travel length:
\begin{equation}\label{}
    P(\Delta r)\propto  \frac{1}{D(\Delta r)}\sim \frac{1}{\Delta r},
\end{equation}
where $D(\Delta r)$ is the physical cost function associated with the travel distance $\Delta r$. Consequently, $\beta\approx1$.

Zone 2 refers to displacements larger than 10 km and shorter than 300 km. These displacements refer mostly to short visits to nearby cities and a typical individual would feel this region as a \emph{medium} zone. In fact, a person can cover such distances up to a few hours with a car or train and round trips within this zone may occur just within the day.
  Within this range the displacement probability distribution is approximately proportional to $1/\Delta r^2$, as depicted in Figure 2. This pattern may emerge because within this displacement range   both fuel cost and time duration costs may be taken in serious consideration by a traveler, who would prefer to save both fuel and time.
  Note that for displacements between 10 km and 300 km both the displacement time duration and fuel costs are not negligible for most people. Thus, for displacements between 10 km and 300 km, that is for trips to nearby cities and towns,  the displacement probability is likely the product of two probability functions which are both proportional to $1/\Delta r$. For these displacements we have
\begin{equation}\label{}
    P(\Delta r)\propto  \frac{1}{D(\Delta r)}\frac{1}{T(\Delta r)}\sim \frac{1}{\Delta r^2},
\end{equation}
where $T(\Delta r)$ is the temporal cost function associated with the displacement.
Consequently, the resulting inverse power law exponent is $\beta\approx2$.

The emergence of the above statistical pattern with an inverse power law exponent with $\beta=2$ may also be favored by a geometrical constrain similar to that expressed in Eq. \ref{pron}. In fact, it may be that each individual is looking for a rare benefit not found in his/her city of residence. An individual may feel that the desired rare benefit is located just in one specific  nearby cities and would a-priory disregard other possible nearby cities as well as most of the country area region among cities, which  would a-priory be considered irrelevant. In this case, while  the probability of \emph{not} finding a desired benefit within the distance $\Delta r<R$ would still decrease with the area of a circle of radius $R$ and should be proportional to $R^{-2}$,  the probability of finding a benefit at a distance $\Delta r=R$ would be likely proportional  only to the size of a specific nearby  city itself and, therefore, it would be independent of the distance $R$ between the two cities. In this specific situation the combined probability between event A and event B is given by the following equation:

\begin{equation}\label{pron2}
    P(A \cup B)=P(A)P(B)\propto \frac{1}{R^2} \times cityarea \propto \frac{1}{R^2},
\end{equation}
which is a Pareto distributions of the displacement length with exponent $\beta=2$. Thus, for travels to nearby cities an agent may assume that the desired benefits are not uniformly distributed within the entire available territory area, but s/he appears to be guided by a pre-decision that limits the search inside a restricted region which is more directly  related to the area of a specific nearby city than to the distance between the residence city and the visited city.

Zone 3 refers to displacements larger than 300 km. These displacements refer mostly to trips to far cities which are likely felt by individuals as a \emph{far} zone because they require multiple day trips. In the eventuality that the data were collected from a phone provider localized in one of the minor European country, let us say Hungary which has an extension of 200-500 km, these far displacements would mostly correspond to abroad trips.
Within this range, Figure 2 shows that the displacement probability distribution may be characterized by an exponent $\beta=3$ and may end with an higher exponent or an exponential cut off. In this situation, the observed pattern may also be due to poor statistics or to the fact that for displacements longer than 300 km some geographical boundary is approached.

However, this third zone may be approximately fit with an equation proportional to $1/\Delta r^3$ suggesting that at least one additional cost mechanism proportional to the displacement length may be activated for these long trips.  For example, in addition to travel fuel and duration costs, a traveler would have lodging costs because s/he will not be able to return home the same day. These additional costs may also increase with the displacement length because a traveler would try to optimize his economical resources by staying longer in a farer location for better exploring it by taking into advantage the rare visit.
 Thus, we can expect that the displacement probability distribution relative to this third displacement zone may decay as an inverse power law with exponent  $\beta = 3$ or a higher order, and may eventually end with an exponential cut off. So,  we have:
\begin{equation}\label{}
    P(\Delta r)\propto  \frac{1}{D(\Delta r)}\frac{1}{T(\Delta r)}\frac{1}{H(\Delta r)}\sim \frac{1}{\Delta r^3},
\end{equation}
 where $H(\Delta r) \approx  \Delta r+O(\Delta r^2)$ is the cost function associated to the additional costs that these long trips require. Consequently, the inverse power law exponent is $\beta \gtrapprox 3$.

Figure 3 depicts a schematic representation of human motion with L\'evy flight displacements in the three different displacement zone ranges. The shadowed circles indicate the urban area of the cities where most displacements occur. Within the shadowed city area the benefits are likely uniformly/randomly distributed and the displacement  distribution obey to a Pareto law with exponent $\beta=1$ as predicted by Eq. \ref{pron}. When people move to nearby cities   the searched benefits are still assumed to be  uniformly/randomly distributed mostly within the shadowed cities zone, and the displacement  distribution obey to a Pareto law with exponent $\beta=2$ as predicted by Eq. \ref{pron2}. When far cities are visited, the  displacement  distribution obey to a Pareto law with exponent $\beta \gtrapprox 3$ because of additional activated  cost mechanisms.

\section{Displacement distributions of walking people: range below 1 km}

An interesting exercise would be to determine whether the above model may somehow be extended to describe and/or predict the behavior of people who just walk below the 1 km region. The data we have analyzed in Sections 2 and 3 do not have such a resolution and refer to displacements from 1 km to 1000 km obtained by means of standard vehicles such as cars, buses, trains and airplanes. Adopting displacements larger than 1 km filters out possible influences due to the small-scale topology of the local geography. Moreover, the fact that standards equivalent transportation vehicles are used by all agents implies that the travel costs both in money and time are evaluated equally by all agents. Consequently, those costs are in first approximation only function of the displacement length and would not significantly depend on the physical characteristic of an individual person. In fact, in these cases, the costs in money and/or time for covering an equal displacement would be approximately the same for a man and a woman, for a young person and for an elderly, for an athletic person and a sedentary one. These common properties would favor the emergence of clearer distribution exponents very close to integer numbers such as $\beta=1$ and $\beta=2$, which would depend only on the displacement range zone, as explained above.

About the human mobility behavior  below the 1 km region, where people more likely walk, we still expect that the probability distributions of their displacements is determined by inverse power laws because of the general theoretical rule discussed in the previous section. In these cases, however, the subjective physical characteristics of the walker, as well as the small-scale topology of the local geography may induce a large variation among the detected exponents. However, because the major cost functions would still depend on one or, at most, two major cost components such as time and physical energy required for each displacement, we would expect that the behavior of walkers below the 1 km region may theoretically mirror at a lower scale the statistical behavior observed at larger scales. Also the geometrical conditions expressed in Eqs. \ref{pron} and \ref{pron2} would be mirrored at a lower scales because people may decide to randomly move only within specific limited area regions where Eq. \ref{pron} works and between these area regions where Eq. \ref{pron2} works. In other words, we would expect that the adoption of vehicles such as cars and trains only magnify the range of human displacement, but does   change the general properties that humans adopt to decide the frequency of their displacements and how they organize the space around them.

 Thus, we do expect that also when people just walk, they still divide the space into three region zones (close, medium, far distances) as depicted in a larger scale in Figure 2, and that the displacement distributions of walking people may be characterized by exponents $\beta \approx1$ at the smallest scale end (for example $\Delta r \lesssim 100-200$ m) and $\beta \approx 2$ at larger scale ($100-200\lesssim \Delta r \lesssim 1000$ m).

 Note that the \emph{far} zone (above about 1 km) will unlikely be covered by simply walking by most people in our days, and distribution of experimental data not appropriately designed to exclude automated vehicles (for example, cars and buses) would not clearly manifest the third zone. In fact, if automated vehicles were used for the covering these larger distances, the statistics implemented in Figure 2 would be activated. Thus, typical distribution of mostly walking people would at most manifest two zones.

The expected 100-200 meter threshold between displacement zone 1 and displacement zone 2, which are characterized by just one or two activated cost mechanisms, respectively,  may be associated to the typical human high resolution visual range area, which limit a natural range around each person, and may have been determined by  evolution. In fact, at smaller scales (less than 100-200 meters) a typical person would be able to see even small details such as small objects and animals. This visual range would constitute a natural close range and would be psychologically equivalent to the residence city space once that fast vehicles are used to cover these larger distances within just a few minutes. In this \emph{close} range zone the decision of moving may be determined by only one cost mechanism and it would be treated as an area where the benefits are randomly/uniformely distributed. This configuration activates the statistics implemented in Eq. \ref{pron}. Thus,  we should expect Pareto distributions with $\beta \approx 1$. On the contrary, for longer scales (larger than 100-200 meters) the displacement has to be more carefully planned because it would involve locations not in the immediate vicinity of the agent (like in the case of visiting nearby cities and towns), and we should expect that the displacement decision is made of two cost mechanisms and the geometrical configuration of displacements between areas of interests activates Eq. \ref{pron2}. Consequently, we should expect Pareto distributions with $\beta \approx 2$.

  The above exponents, $\beta=1$ and $\beta=2$ respectively, should be expected in the cases in which the individual movement is not strongly influenced by specific environmental topology. In fact, when people are forced by  environmental topology to cover larger distances more often and small distanced less often, we would expect approximate  Pareto distributions with $1<\beta <2$, where the threshold between  zone 1 and zone 2 becomes less evident.

The above predictions appear to be fully compatible with the recent results of Rhee et al. \cite{Rhee} (see their figures 5 and 6), who studied the Levy-walk nature of human mobility of mostly walking people based on 226 daily GPS traces collected from 101 volunteers in five different outdoor sites such as two university campuses, in Manhattan (NY), in Disney World in Orlando (FL), and one state fair.

Rhee et al. \cite{Rhee} used truncated Pareto law to fit his distributions which may present the same problem of occulting the real physics of the process.
Interestingly, among the five records studied by Rhee et al.,  the one that detected an inverse power law exponent closer to 1 ($\beta=1.16$ within the range 0.1-10 km), which would correspond to our finding of figure 2, was the one relative to Manhattan (NY) where the displacement means included vehicles such as cars and buses, as for the data used for obtaining the distribution depicted in Figures 1 and 2. Movements inside a city also more likely activates the statistics implemented in Eq. \ref{pron}. Thus,  we should have expected a Pareto distribution with $\beta \approx 1$.

We observe that the walking in the two university campuses (North Carolina State University $\beta =1.29$; Korea Advanced Institute of Science and Technology $\beta=1.66$) could be strongly constrained by geographical topology because students are more likely forced to move to specific locations, and here we found Pareto distributions with $1<\beta <2$, where the threshold between zone 1 and zone 2 is not evident.

On the contrary, the walking in the Disney World ($\beta=1.82$) and the state fair in Raleigh (NC) ($\beta=1.74$) is likely less constrained by specific geographical topological necessities, and these data may more accurately describe general properties of free human mobility. However, it is an impression of the present author that the distributions depicted in figure 6d and 6e in Rhee et al. \cite{Rhee} referring to the latter two cases would be better fit with a two-scale Pareto law, and present a threshold from a low Pareto exponent closer to $\beta=1$ to a high Pareto exponent closer to $\beta=2$ around 100-200 m. This claim is proven herein in Figure 4 that reproduces figure 6d in Rhee et al., where the pattern is clearer because apparently obtained with more data.

The finding depicted in Figure 4 appears to confirm our expectation that the behavior of simple walkers below the 1 km region  approximately mirror at a lower scale the behavior observed at larger scales depicted in Figure 2. We see two displacement zones characterized by $\beta=1$ and $\beta=2$, respectively, where one or two cost mechanisms are likely taken into account by a walker according to whether the displacement is felt as close or medium. Moreover, we notice that Disney World park is organized as an agglomerate of little town/parks that resemble in a miniaturized scale the typical geographical topology of a set of close urban areas.  Thus, it may be not surprising that the displacement distribution depicted in Figure 4 looks similar to that depicted in Figure 2 by reproducing at a smaller scale the geometrical/statistical properties of Zone 1 and Zone 2 that activate the conditions expressed in Eqs. \ref{pron} and \ref{pron2}, which yield to Pareto distributions with $\beta=1$ and $\beta=2$, respectively.

\section{Conclusion}

In this work we have studied the distribution depicted in Figure 1 that is made of a sample of several million of human displacement data \cite{Gonzalez}. These authors have traditionally fit their distribution with a typical exponentially truncated inverse power law with one exponent without being able to explain its physical origin nor the physical meaning of the measured exponent.

However, it is legitimate to look for alternative ways of fitting a complex probability distribution \cite{Beck} if this exercise yields to a physical explanation of the phenomenon under study. We have found that the same distribution may be fit with a multi-scale inverse power law distribution with simple integer exponents 1, 2 and 3 within three zones that depend on the length of the displacement: from  1 to 10 km, from  10 to  300 km and from 300 to 1000 km, respectively.

The multiple integer exponent distribution may be generated by the fact that people optimize their movements by first dividing the entire available space into three zones corresponding to close, medium and far distances. Then they assume that on average the \emph{benefits} of travel are distributed on average randomly and uniformly within specific area of interests such as cities at all travel distances. Finally, people would take into consideration a incremental number of cost mechanisms according to each zone, which are always approximately proportional to the displacement length such as travel fuel cost, travel time duration and others.  Thus, as we have proven,   each single mechanism alone would yield to a Pareto probability distribution with exponent $\beta=1$ because this kind of distribution would fit a geometrical constrain related to the condition of benefits randomly and uniformly within specific area of interests as expressed in Eq. \ref{pron} that yields to Pareto distribution with $\beta=1$. The combination of two or more cost mechanisms would yield to Pareto probability distribution with exponents $\beta$ = 1, 2, 3 and so on, according to the displacement zone that activates 1, 2, 3 (and so on) cost mechanisms. In particular, trips to nearby cities would be constrained by a geometrical condition expressed in Eq. \ref{pron2} that yields to Pareto distribution with $\beta=2$. This geometrical constrain requires that when a person search a benefit in a nearby city s/he would still search it within a specific city/area like region, and s/he would a-priory disregard any other region at equivalent distance in other directions.

The above conclusion appears banal and predictable. Everybody knows that when people travel they assume that specific benefits may be randomly/uniformly distributed in areas of interests such as within cities, and take into account things such as fuel cost, time duration etc. It is perfectly reasonable that a-priory, on average, individuals divide the space in three major zones (close, medium, far distances). Thus, because of simple geometrical constrains we should have expected that human displacements could be described by simple inverse power law distributions with simple integer exponents.  However, the importance of our work is to have proved this fact for the first time by using an actual human mobility distribution.

In Section 5 we have show that the same model is capable to predict and explain also   the complexity of the L\'evy-walk nature of human mobility for mostly walking people, as Figure 4 shows. In fact, the adoption of vehicles simply magnifies the geographical space taken into account by an individual, but the physical mechanisms remain the same. Walking people, too, would first divide the space into areas of interests and three potential regions (close, medium and far distances), would assume that in each area the benefits are homogeneously distributed and would associate to each region an increasing integer number of cost mechanisms. This yields to multi-scale Pareto distributions with incremental exponents $\beta$ = 1, 2 (and so on). However, the third region may not appear in current experimental distributions for the reasons explained in the text.

The multi-scale mobility model described in Eqs. \ref{eq2}, \ref{eq3}, \ref{pron} and \ref{pron2} can be quite general, indeed, and can efficiently explain several real mobility and diffusion processes where the displacement is reasonably a function of geometrical constrains and the number of cost functions involved in the process.
This cost function geometrically increases with the displacement length every time that a certain displacement zone is crossed.  This yields to apparently truncated power-law distributions that, indeed, are made of different power law distributions with incremental exponents (in the human mobility case $\beta =   1, 2, 3, \dots$) referring to different displacement length ranges, as suggested by the data depicted in Figures 2 and 4.

\newpage

\begin{figure}
\includegraphics[angle=-90,width=30pc]{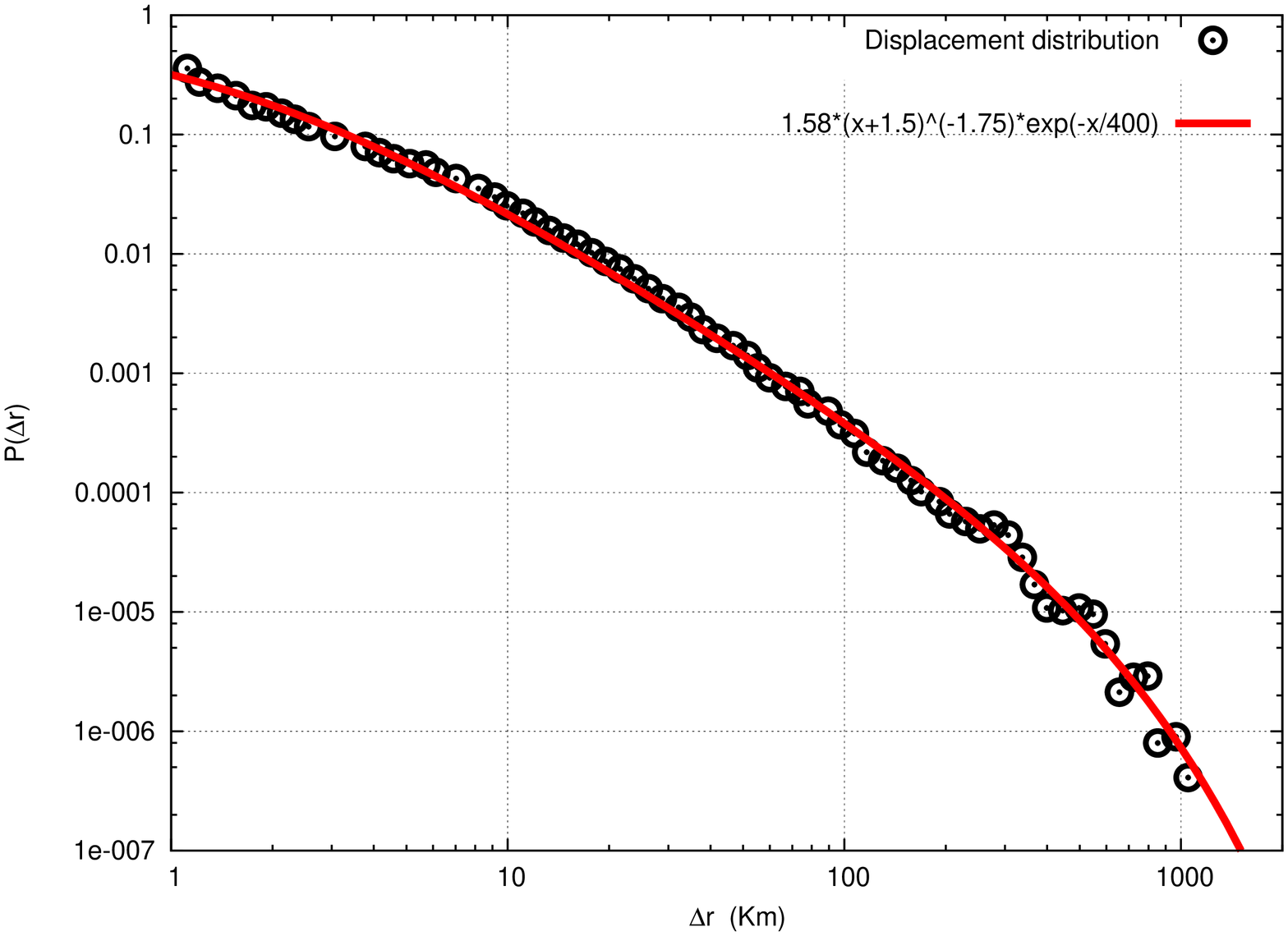}
\caption{ Distribution of human displacements (from record D1 and Figure 1C in Gonzalez \emph{et al.} \cite{Gonzalez}). The length $\Delta r$ is the travel distance.  The distribution is fit with Eq. (\ref{eq1}).   }
\end{figure}

\begin{figure}
\includegraphics[angle=-90,width=30pc]{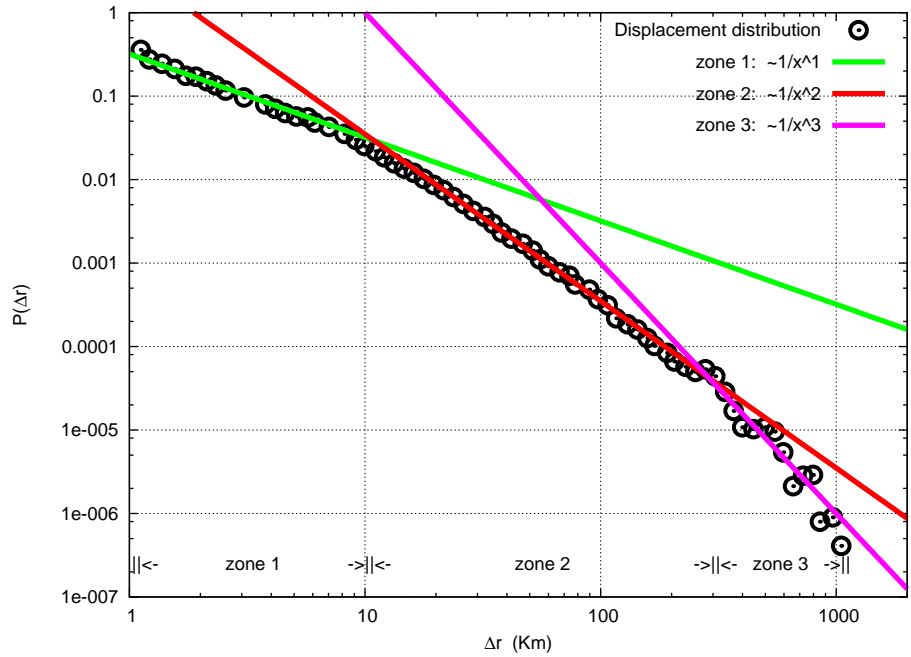}
\caption{ Distribution of human displacements (from record D1 and Figure 1C in Gonzalez \emph{et al.} \cite{Gonzalez}). The length $\Delta r$ is the travel distance.  The distribution is approximately reproduced within three consecutive ranges with incremental inverse power law functions proportional to $1/\Delta r^\beta$. We get: 1) $\beta  \approx 1$ for  1 km $\lesssim\Delta r \lesssim $ 10 km;  2)  $\beta  \approx 2$ for  10 km $\lesssim \Delta r \lesssim$ 300 km;   3) $\beta  \approx 3$ for  300 km $\lesssim \Delta r \lesssim $ 1000 km.  }
\end{figure}

\begin{figure}
\includegraphics[angle=0,width=35pc]{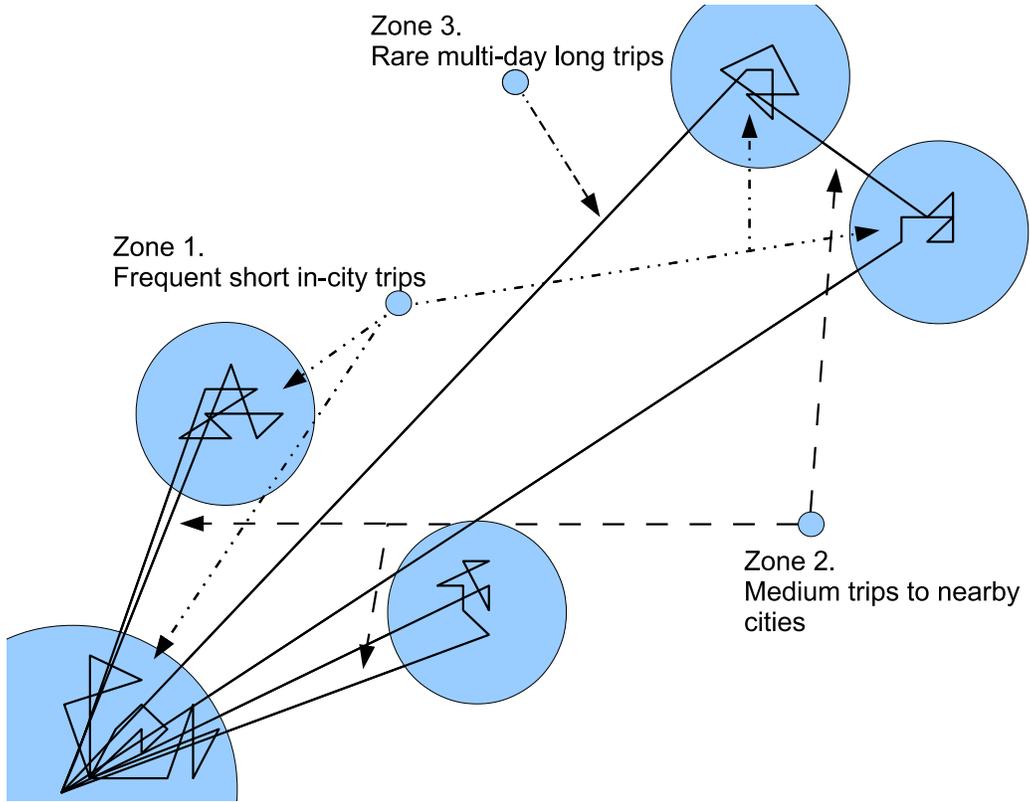}
\caption{ Schematic representation of human motion with L\'evy flights in three different displacement zone ranges. The bottom left corner represents the residence location where the walker returns. The regions with a higher density of displacements represent cities or towns (zone 1) where the walker stays longer. The distance among the cities classifies zone 2 or zone 3 according to whether the journey is medium or long.  The shadowed areas represent the urban area where the benefits are assumed to be randomly/uniformly distributed. Displacements within these restricted areas  yield to Pareto distribution with exponent $\beta=1$ according to Eq. \ref{pron}. Displacements between close  areas of interest  yield to Pareto distribution with exponent $\beta=2$ according to Eq. \ref{pron2}.}
\end{figure}

\begin{figure}
\includegraphics[angle=0,width=30pc]{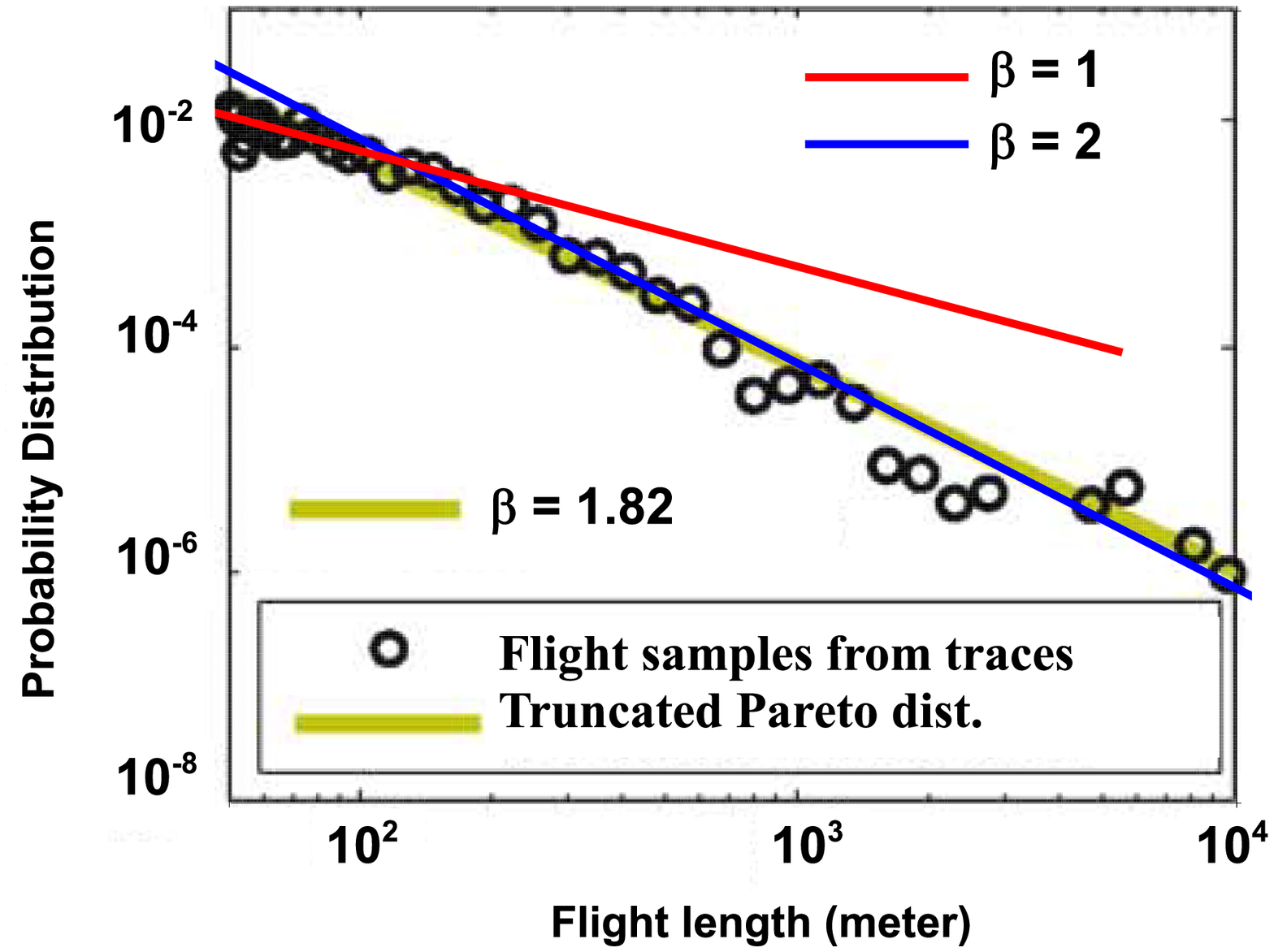}
\caption{Reproduction of figure 6 in Rhee et al. \cite{Rhee} which refers to a distribution of displacements at Disney World (Orlando, FL) by a sample of traces obtained from 18 volunteers that mainly walked in the park during holidays. The figure shows in green the authors' original fit obtained with a truncated Pareto distribution that gives an exponent $\beta=1.86$ for the entire range. The red and blue curves, instead, were added by me and highlight that the smallest displacement range up to 100-200 meter is compatible with a Pareto distribution with $\beta = 1$, while the longer scale is compatible with a Pareto distribution with $\beta = 2$. Note that the statistics for displacements above approximately 1000 m is visibly poorer.}
\end{figure}


\begin{thebibliography}{99}


\bibitem{Scafetta2010} N. Scafetta, Fractal and Diffusion Entropy Analysis of Time Series: Theory, concepts, applications and computer codes for studying fractal noises and Lévy walk signals.  (VDM Verlag Dr. Müller 2010).

\bibitem{Scafetta} N. Scafetta, D. Marchi and B. J. West, ``Understanding the complexity of human gait dynamics,'' Chaos \textbf{19}, 026108 (2009).

\bibitem{Horner} M. W.	Horner, and M. E. S O'Kelly, ``Embedding economies of scale concepts for hub networks design,'' J. Transp. Geogr. \textbf{9}, 255 (2001).

\bibitem{Kitamura1}  R. Kitamura, S. Fujii and E. Pas, ``Time-use data, analysis and modeling: toward the next generation of transportation planning methodologies,'' Transport Policy \textbf{4}, 225 (1997)..

\bibitem{Kitamura2}    R. Kitamura, C. Chen, R. M. Pendyala, and R. Narayaran, ``Micro-simulation of daily activity-travel patterns for travel demand forecasting,'' Transportation \textbf{27}, 25 (2000).


\bibitem{Colizza} 	V. Colizza, A. Barrat, M. Barthelemy, A.-J. Valleron,  and A. Vespignani,  ``Modeling the worldwide spread of pandemic influenza: baseline case and containment interventions,'' PLoS Medicine \textbf{4}, 95 (2007).

\bibitem{Eubank} S.	Eubank et al., ``Controlling epidemics in realistic urban social networks,'' Nature \textbf{429}, 180 (2004).

\bibitem{Hufnagel} L.	Hufnagel, D. Brockmann, and T. Geisel, ``Forecast and control of epidemics in a globalized world,'' Proc. Natl Acad. Sci. USA \textbf{101}, 15124 (2004).

\bibitem{Kleinberg} J. Kleinberg, ``The wireless epidemic,'' Nature \textbf{449}, 287 (2007).

\bibitem{Schnettler}  S. Schnettler,	``A structured overview of 50 years of small-world research,''
Social Networks \textbf{31}, 165 (2009).


\bibitem{Toivonen} R. Toivonen, L. Kovanen, M. Kivel\"a, J. Onnela, J. Saram\"aki, and K. Kaski, ``A comparative study of social network models: Network evolution models and nodal attribute models,'' Social Networks \textbf{31},  240 (2009).

\bibitem{Gonzalez} M. C.	Gonzalez, C. A. Hidalgo, and A.-L. Barabasi, ``Understanding individual human mobility patterns,'' Nature \textbf{453}, 779 (2008).

\bibitem{Brockmann} D. D.	Brockmann, L. Hufnagel, and T. Geisel, ``The scaling laws of human travel,'' Nature \textbf{439}, 462 (2006).




\bibitem{Viswanathan}  G. M. Viswanathan, \emph{et al. }, ``L\'evy  flight search patterns of wandering albatrosses,''
Nature \textbf{381}, 413 (1996).


\bibitem{Edwards}   A. M.	Edwards et al., ``Revisiting  L\'evy  flight search patterns of wandering albatrosses, bumblebees and deer,'' Nature \textbf{449}, 1044 (2007).

\bibitem{Ramos} G.	Ramos-Fernandez et al., ``L\'evy  walk patterns in the foraging movements of spider monkeys (Ateles geoffroyi),'' Behav. Ecol. Sociobiol. \textbf{273}, 1743 (2004).

\bibitem{Sims} D. W.	Sims et al., ``Scaling laws of marine predator search behaviour,'' Nature \textbf{451}, 1098 (2008).





\bibitem{Rhee} I. Rhee, M. Shin, S. Hong, K. Lee, S.J. Kim, and S. Chong, ``On the Levy-Walk Nature of Human Mobility,'' IEEE-ACM Transactions on Networking \textbf{19:3},   630 (2011).



\bibitem{Watts}  D. J. Watts, Six Degrees: The Science of a Connected Age. (Norton \& Company 2003).

\bibitem{Schroeder} M. Schroeder, Fractals, chaos, power laws: minutes from an infinite paradise. (Freeman, New York 1991).

\bibitem{Latapy} M. Latapy, C. Magnien, and  N. Del Vecchio, ``Basic notions for the analysis of large two-mode networks,'' Social Networks \textbf{30},  31 (2008).

\bibitem{Beck} C.	Beck and E. G. D. Cohen,  ``Superstatistics,''  Physica A \textbf{322}, 267 (2003).

\bibitem{Fenner}  	 T. Fenner, M. Levene, and G. Loizou, ``A model for collaboration networks giving rise to a power-law distribution with an exponential cutoff,'' Social Networks \textbf{29},  70 (2007).

\bibitem{Lundevaller} E. H. Lundevaller, ``The effect of travel cost on frequencies of shopping and recreational trips in Sweden,'' J. of Transport Geography \textbf{17},  208 (2009).

\bibitem{Fujita}  M. Fujita, P. Krugman, and A. J. Venables, The Spatial Economy: Cities, Regions, and International Trade, The MIT Press (September 1, 2001).

\bibitem{Mokhtarian}  P. Mokhtarian, and C. Chen, ``Ttb or not ttb, that is the question: a review and analysis of the empirical literature on travel time (and money) budgets,'' Transportation Research Part A \textbf{38}, 643 (2004).

\end{thebibliography}
\end{document}